\documentclass[%
 twocolumn, amsmath,amssymb, aps]{revtex4}
\pdfoutput=1
\usepackage{graphicx}
\usepackage{dcolumn}
\usepackage{bm}

\begin{document}

\preprint{APS/123-QED}

\title{Role of electric charge in shaping equilibrium configurations of fluid
tori encircling black holes}

\author{Ji\v{r}\'{i} Kov\'{a}\v{r}}
  \email{Jiri.Kovar@fpf.slu.cz}
\author{Petr Slan\'{y}}
\author{Zden\v{e}k Stuchl\'{i}k}%
\affiliation{Institute of Physics, Faculty of Philosophy and Science,
Silesian University in Opava\\ Bezru\v{c}ovo n\'{a}m. 13, CZ-74601 Opava, Czech Republic}%

\author{Vladim\'{i}r Karas}
\affiliation{Astronomical Institute, Academy of Sciences,
Bo\v{c}n\'{i} II, CZ-14131\,Prague, Czech Republic}

\author{Claudio Cremaschini}
\affiliation{SISSA \& INFN, Via Bonomea 265, I-34136 Trieste, Italy}

\author{John C. Miller}
\affiliation{SISSA \& INFN, Via Bonomea 265, I-34136 Trieste, Italy}
\affiliation{Department of Physics (Astrophysics), University of Oxford\\
Keble Road, Oxford OX1 3RH, U.K.}

\begin{abstract}
 Astrophysical fluids may acquire non-zero electrical charge because of strong 
irradiation or charge separation in a magnetic field. In this case, 
electromagnetic and gravitational forces may act together and produce new 
equilibrium configurations, which are different from the uncharged ones. 
Following our previous studies of charged test particles and uncharged perfect 
fluid tori encircling compact objects, we introduce here a simple test model of 
a charged perfect fluid torus in strong gravitational and electromagnetic 
fields. In contrast to ideal magnetohydrodynamic models, we consider here the 
opposite limit of negligible conductivity, where the charges are tied 
completely to the moving matter. This is an extreme limiting case which can 
provide a useful reference against which to compare subsequent more complicated 
astrophysically-motivated calculations. To clearly demonstrate the features of 
our model, we construct three-dimensional axisymmetric charged toroidal 
configurations around Reissner-Nordstr\"{o}m black holes and compare them with 
equivalent configurations of electrically neutral tori.
\end{abstract}

\maketitle


\section{Introduction}
 Equilibrium toroidal configurations of perfect fluid play an important role in 
studies of geometrically thick accretion discs around compact objects 
\cite{Fra-Kin-Rai:2002:AccretionPower:}. The isobaric surfaces also have 
toroidal topology and in order for accretion to occur there must be a critical, 
marginally-closed isobaric surface with a cusp through which matter can outflow 
from the disc onto the compact object. In the following, we focus on black-hole 
systems and ignore self-gravity of the disc material. Shapes and properties of 
the tori, such as pressure and density profiles, are then determined by the 
black-hole spacetime geometry, an appropriately chosen rotation law (giving the 
distribution of specific angular momentum), and the fluid parameters.

Perfect fluid tori in Schwarzschild and Kerr backgrounds were extensively 
discussed in the original fundamental papers establishing this line of work 
\cite{Koz-Jar-Abr:1978:ASTRA:,Abr-Jar-Sik:1978:ASTRA:}. Later on, many studies 
appeared generalizing these models and including further details 
\cite{Stu-Sla-Hle:2000:ASTRA:,Fon-Dai:2002:MNRAS:,Rez-Zan-Fon:2003:ASTRA:,Stu:2005:MODPLA:,Sla-Stu:2005:CLAQG:,Stu-Sla-Kov:2009:CLAQG:,Kuc-Sla-Stu:2011:CASTRP:}, 
describing tori also in Schwarzschild-de Sitter, Kerr-de Sitter and 
Reissner-Nordstr\"{o}m-de Sitter spacetimes, where presence of the so-called 
static radius \cite{Stu-Hle:1999} implies also the existence of tori with an 
outer cusp.

The material in accretion discs contains charged particles (which may or may 
not be quasi-neutral in bulk) and the central black hole might also be charged. 
The charged or quasi-neutral fluid creates its own electromagnetic field which 
would then couple with that of the black hole, leading to a different and much 
more complicated description of the motion. Using the equations for the 
dynamics of the fluid and specifying its `internal' properties (conductivity, 
viscosity, equation of state, etc), one can solve the system so as to obtain 
profiles for the four-velocity, pressure, matter density and charge density 
\cite{Pun:2008}. However, the system of equations is rather complex, and in 
general requires the use of sophisticated numerical approaches and codes, even 
if simplifying assumptions are made such as taking infinite electrical 
conductivity (the limit of `ideal magnetohydrodynamics'), no self-gravity, etc. 
On the other hand, some characteristic features of the motion of quasi-neutral 
or charged fluid, can also be studied relatively simply in a semi-analytic way 
\cite{Lov-etal:1986,Pra-etal:1989,Tri-etal:1990,Bha-Pra:1990,Bha-etal:1990,Ban-etal:1995,Ban-etal:1997,Kom:2006}.

The approximation of ideal magnetohydrodynamics (MHD) is reasonable in many 
astrophysically-relevant situations involving fluids in motion \cite{Mel:1980}. 
However, there are other physical circumstances in which it is important to 
include the effects of finite conductivity \cite{Koi:2010,Pal-etal:2009,Kud-Kab:1996}, and there the behavior becomes 
more complex, especially when strong gravitational and external magnetic fields 
are also present. In order to address some of these effects in their mutual 
interplay, it can be useful to look also at the opposite limit to that of ideal 
MHD: the limit of negligibly small conductivity.

Here, we examine the problem of the interaction between charged moving 
matter and the gravitational and electrostatic fields of the black hole, 
concentrating on an idealized situation which allows us to illustrate some 
otherwise very complicated effects. We consider a simple test model in which 
the matter is taken to be slightly charged and electrically non-conductive 
(dielectric), with the aim of providing an extreme reference case against which 
to compare subsequent more detailed calculations. We proceed by first 
specifying a prescribed form for the angular momentum distribution, and then 
solving the dynamical equations to find the profiles of pressure, mass density 
and charge density. This approach can be seen as generalizing the studies of 
uncharged perfect-fluid tori mentioned above by adding the charge. Note that, 
throughout, our tori are assumed to be composed of test fluids in which both 
the self-gravitational and self-electromagnetic fields are neglected. This 
gives a useful simplification, which is acceptable for weakly-charged, low-mass 
tori that have hardly any effect on the spacetime geometry or the ambient 
electromagnetic field. This approach helps us in building a semi-analytic 
model. In principle, inclusion of self-gravity of the tori (following, 
e.g. \cite{Nis-Eri:1994,Kar-Hur-Sem:2004,Ste:2011}) and self-electromagnetic effects (see, e.g. 
\cite{Bha-Pra:1990}) would be possible, but that would enormously complicate 
the situation.

The charged dielectric perfect-fluid tori can also be seen as generalizing 
studies of charged test particles orbiting around black holes 
\cite{Cal-Fel-Fab-Tur:1982,Pra-Sen:1994,Vok-Kar:1991a,Bic-Stu-Bal:1989,Bal-Bic-Stu:1989,Stu-Bic-Bal:1999,Fel-Sor:2003,Stu-Hle:1998,Bak-Sra-Stu-Tor:2010}. Various aspects of the charged test particle motion were also discussed in 
\cite{Pre:2010,Kop-Kar-Kov-Stu:2010}, concerning the possibility of collimated 
ejection along the axis of a rotating magnetized black hole, while 
investigation of stable off-equatorial lobes of charged particles was discussed 
in \cite{Kov-Stu-Kar:2008,Stu-Kov-Kar:2009,Kov-Kop-Kar-Stu:2010}. In general, 
the motion of test particles is bounded within effective potential wells and 
this represents a model for a very dilute toroidal structure consisting of 
non-interacting particles; here, we add the non-electrical interaction between 
them, the pressure. Commonly, pressure leads to geometrically thick 
structures extending further out of the equatorial plane.

In section II, we present the basic equations. In section III, we apply them to 
the case of a torus around a charged, non-rotating black hole described by the 
Reissner-Nordstr\"{o}m metric. Nevertheless, the presented approach is 
suitable for a description of charged tori near to any models of compact 
objects with well-defined geometry and electromagnetic field. We chose the 
Reissner-Nordstr\"{o}m black hole because of its extremely clear external 
electric field and geometry, given in an analytic form. For illustrative purposes we set the charge of the central black hole
considerably exceeding astrophysically realistic values; our paper presents a toy model exhibiting the physical mechanism of mutual
interaction between charged fluid and a black hole.
In section IV, we discuss the form of the isobaric surfaces for a torus with constant specific 
angular momentum composed of an uncharged barotropic perfect fluid. This is 
then extended to charged tori in section V, where we also present a comparison 
between equivalent charged and uncharged cases. 
This work involves making a number of simplifying assumptions, and we discuss the nature and impact of 
these (including the zero-conductivity limit) in section VI. Section VII is the 
conclusion. Throughout the paper, we use the geometrical system of units 
($c=G=1$) and metric signature $+2$.

\section{Basic equations}
 In general, the motion of charged perfect fluid is described by two sets of 
general relativistic MHD equations. These are the conservation laws and 
Maxwell's equations
 \begin{eqnarray}
\label{cons}
\nabla_{\beta}T^{\alpha\beta}&=&0,\\
\label{Maxw}
\nabla_{\beta}F^{\alpha\beta}&=&4\pi J^{\alpha},
\end{eqnarray}
 where the \mbox{4-current} density $J^{\alpha}$, which satisfies the 
continuity equation
 \begin{eqnarray}
\nabla_{\alpha}J^{\alpha}=0,
\end{eqnarray}
 can be expressed in terms of the charge density $q$, conductivity $\sigma$ and 
\mbox{4-velocity} $U^{\alpha}$ of the fluid by using Ohm's law
\begin{eqnarray}
\label{Ohm}
J^{\alpha}=q U^{\alpha}+\sigma F^{\alpha\beta}U_{\beta},
\end{eqnarray}
 with the electromagnetic tensor $F^{\alpha\beta}$ being given in terms of the 
vector potential by 
$F_{\alpha\beta}=\nabla_{\alpha}A_{\beta}-\nabla_{\beta}A_{\alpha}$. This 
electromagnetic tensor describes the vacuum external electromagnetic field of 
the compact object (which pervades the fluid), and also the internal 
electromagnetic field of the fluid itself, i.e.,
 \begin{eqnarray}
F^{\alpha\beta}=F^{\alpha\beta}_{\rm EXT}+F^{\alpha\beta}_{\rm INT}.
\end{eqnarray}
The stress-energy tensor $T^{\alpha\beta}$ consists of matter and 
electromagnetic parts
 \begin{eqnarray}
T^{\alpha\beta}=T^{\alpha\beta}_{\rm MAT}+T^{\alpha\beta}_{\rm EM},
\end{eqnarray}
where
\begin{eqnarray}
\label{Tmat}
T^{\alpha\beta}_{\rm MAT}&=&(\epsilon+p)U^{\alpha}U^{\beta}+pg^{\alpha\beta},\\
\label{Tem}
T^{\alpha\beta}_{\rm EM}&=&\frac{1}{4\pi}\left(F^{\alpha}_{\;\;\gamma}F^{\beta\gamma}-\frac{1}{4}F_{\gamma\delta}F^{\gamma\delta}g^{\alpha\beta}\right).
\end{eqnarray}
 Besides the pressure $p$ and energy density $\epsilon$, the other fluid 
variables are the rest-mass density $\rho$ and the specific internal energy 
$\varepsilon=\epsilon/\rho-1$. The thermodynamical description of the fluid is 
specified by supplying an appropriate equation of state $p=p(\epsilon,q)$, 
which also involves the charge density of the fluid, describing the 
contribution of the Coulomb interaction between the fluid particles to the 
total pressure.

We build our model by considering a non-conductive ($\sigma=0$) charged test 
fluid (taken to be a perfect fluid) in an axially symmetric spacetime and 
use spherical polar coordinates $(t,r,\theta,\phi)$. The fluid rotates in the 
\mbox{$\phi$-direction} with \mbox{4-velocity} $U^{\alpha}=(U^t,U^{\phi},0,0)$, 
specific angular momentum $\ell=-U_{\phi}/U_t$ and angular velocity (related to 
distant observers) $\Omega=U^{\phi}/U^t$, related by the formulae
 \begin{eqnarray}
\label{Omega}
\Omega&=&-\frac{\ell g_{tt}+g_{t\phi}}{\ell g_{t\phi}+g_{\phi\phi}},\\
(U_t)^2&=&\frac{g_{t\phi}^2-g_{tt}g_{\phi\phi}}{\ell^2 g_{tt}+2\ell g_{t\phi}+g_{\phi\phi}}
\end{eqnarray}
 By writing out the covariant derivative of the electromagnetic part of 
the stress-energy tensor (\ref{Tem}) appearing the left hand side of the 
conservation law (\ref{cons}), moving it to the right hand side, and using 
the Maxwell equations (\ref{Maxw}), with 
\begin{eqnarray}
\nabla_{\beta}F^{\alpha\beta}_{\rm EXT}=0,\quad 
\nabla_{\beta}F^{\alpha\beta}_{\rm INT}=4\pi J^{\alpha},
\end{eqnarray}
 we obtain the equation $\nabla_{\beta}T^{\alpha\beta}_{\rm 
MAT}=F^{\alpha\beta}J_{\beta}$, with $J_{\beta}$ being the \mbox{4-current} 
density due to the motion of the charged fluid torus. We are not here including 
the effects of the electromagnetic field generated by this \mbox{4-current}: 
our tori are considered as being composed of `test matter' from the 
electromagnetic point of view as well as from the gravitational one, i.e., 
$F^{\alpha\beta}_{\rm INT}\ll F^{\alpha\beta}_{\rm EXT}$ and we can write 
$F^{\alpha\beta}=F^{\alpha\beta}_{\rm EXT}$. Then we get the master equation
 \begin{eqnarray}
\label{master}
\nabla_{\beta}T^{\alpha\beta}_{\rm MAT}=F^{\alpha\beta}_{\rm EXT}J_{\beta}.
\end{eqnarray}
 Note that in this approach we do not need to solve Maxwell's equations 
(\ref{Maxw}), since the electromagnetic field is prescribed. Also, because of 
the non-conductivity, we have
 \begin{eqnarray}
J^{\alpha}=q U^{\alpha}.
\end{eqnarray}

 The equations of motion (\ref{master}) give two non-linear partial 
differential equations for the pressure $p$, whose profiles we are wanting to 
find:
 \begin{eqnarray}
\label{pressure}
\partial_r p&=&-(\epsilon+p)\left(\partial_r\,\ln{(U_t)} - \frac{\Omega \partial_r \ell}{1-\Omega \ell}\right)-q\, F_{r\alpha}U^{\alpha}\equiv \mathcal{R},\nonumber\\
\partial_{\theta} p&=&-(\epsilon+p)\left(\partial_{\theta}\,\ln{(U_t)} - \frac{\Omega \partial_{\theta} \ell}{1-\Omega \ell}\right)-{q}\, F_{\theta\alpha}U^{\alpha}\equiv \mathcal{T},\nonumber\\
\end{eqnarray}
where $\mathcal{R}=\mathcal{R}(r,\theta)$ and $\mathcal{T}=\mathcal{T}(r,\theta)$.
These equations are not integrable unless the integrability condition
\begin{eqnarray}
\label{condition}
\partial_{\theta} \mathcal{R}=\partial_r \mathcal{T}
\end{eqnarray}
is satisfied, and this is therefore a requirement.  

The existence of a solution is guaranteed if $q=0$, so that the last terms in 
equations (\ref{pressure}) vanish and we get the Euler equation describing a 
rotating uncharged perfect fluid (see, e.g., papers 
\cite{Koz-Jar-Abr:1978:ASTRA:,Abr-Jar-Sik:1978:ASTRA:}). In this case, when the 
angular momentum distribution $\ell=\ell(r,\theta)$ is chosen, a solution of 
the Euler equation for any barotropic fluid (having $p=p(\epsilon)$) can be 
derived from Boyer's condition 
\cite{Koz-Jar-Abr:1978:ASTRA:,Abr-Jar-Sik:1978:ASTRA:}
 \begin{eqnarray} 
\label{Boyer}                 
\int_0^p \frac{{\rm d}p}{p+\epsilon}&=&W_{\rm in}-W,\\             
W_{\rm in}-W&=&\ln{(U_t)}_{\rm in}-\ln{(U_t)}+\int_{\ell_{\rm in}}^{\ell}\frac{\Omega {\rm d}\ell}{1-\Omega\ell},
\end{eqnarray}
 where the subscript `in' refers to the inner edge of the torus in the 
equatorial plane. This condition enables us to straightforwardly determine the 
isobaric surfaces in the torus in terms of the equipotential surfaces of the 
`gravito-centrifugal' potential $W(r,\theta)$: for equilibrium toroidal 
configurations composed of barotropic perfect fluid, the equipotential surfaces 
of $W$ correspond to surfaces of constant pressure (or energy density) in the 
fluid.

When $q\neq 0$, the situation is more complicated and equations 
(\ref{pressure}) are no longer integrable for arbitrary $q(r,\theta)={\rm 
const}$ and arbitrarily chosen $\ell=\ell(r,\theta)$. It is then necessary 
either to specify $\ell=\ell(r,\theta)$ (with even $\ell={\rm const}$ being 
possible) and find an appropriate $q=q(r,\theta)$ which is consistent with that 
or, vice versa, to specify $q=q(r,\theta)$ (with even $q={\rm const}$ being 
possible) and find an appropriate $\ell=\ell(r,\theta)$. However, this is 
strictly `necessary' only if the equation of state is prescribed. Otherwise, 
one could also absorb the constraint into that. The charged tori must clearly 
have distributions of charge and angular momentum which satisfy the 
integrability condition.

\section{Isobaric surfaces in Reissner-Nordstr\"{o}m geometry}
 The Reissner-Nordstr\"{o}m metric, representing the space-time outside a 
charged, non-rotating black hole, provides a suitable mathematically-simple 
test example for experimenting with ideas before moving on to more complicated 
examples having direct astrophysical relevance. In the dimensionless ($M=1$) 
Schwarzschild coordinates, the only free parameter in the line element of the 
Reissner-Nordstr\"{o}m geometry
 \begin{eqnarray}
{\rm d}s^2&=&-\left(1-\frac{2}{r}+\frac{Q^2}{r^2}\right){\rm d}t^2+\left(1-\frac{2}{r}+\frac{Q^2}{r^2}\right)^{-1}{\rm d}r^2\nonumber\\ 
&&+\,r^2({\rm d}\theta^2+\sin^2{\theta}{\rm d}\phi^2)
\end{eqnarray}
 is the dimensionless charge $Q$, which takes values $|Q|\leq 1$ for black-hole 
spacetimes and $|Q|>1$ for naked-singularity spacetimes. The locations of event 
horizons, corresponding to the pseudo-singularities of the geometry, are given 
by solutions of the equation $\Delta\equiv r^2-2r+Q^2=0$. The ambient electric 
field is static and spherically symmetric, like the space-time, and is 
described by the vector potential $A_{\alpha}=(A_t,0,0,0)$, with the non-zero 
component being given by
 \begin{eqnarray}
A_t=-\frac{Q}{r}.
\end{eqnarray}
In this background, the pressure equations (\ref{pressure}) reduce to the form
 \begin{eqnarray}
\label{pressure_RN}
\partial_r p&=&-(p+\epsilon)\left(\partial_r\,\ln{|U_t|} - \frac{\Omega \partial_r \ell}{1-\Omega \ell}\right)+q\, U^t\partial_r A_t,\nonumber\\
\partial_{\theta} p&=&-(p+\epsilon)\left(\partial_{\theta}\,\ln{|U_t|} - \frac{\Omega \partial_{\theta} \ell}{1-\Omega \ell}\right),
\end{eqnarray}
where 
\begin{eqnarray}
U_t=-\frac{r\sin{\theta}\sqrt{\Delta}}{\sqrt{r^4\sin^2{\theta}-\ell^2\Delta}}.
\end{eqnarray}

To proceed further, it is then necessary to choose an equation of state. For an 
uncharged perfect fluid, a suitable choice is to have a polytropic relation 
between the pressure and the rest-mass density
 \begin{eqnarray}
\label{EOS}
p=\kappa \rho^{\Gamma},
\end{eqnarray}
 with $\kappa$ and $\Gamma$ being a polytropic constant and index. This 
widely-used relation is a convenient simple form which embodies conservation of 
entropy (as appropriate for a perfect fluid). For our calculations, we have 
used $\kappa=10^{12}$ and $\Gamma=2$. This value of $\Gamma$ is 
mathematically convenient for making analytic integrations and, while it is 
rather high for physical applications, the convenience makes its use consistent 
within the spirit of the present simple model. We have chosen a high value of 
$\kappa$, because electrostatic corrections to the equation of state then 
become negligible, so that we can continue to use this polytropic equation of 
state consistently even in the charged case (we comment further on this in 
section VI C). Moreover, we use values of $\rho$ which are sufficiently low so 
that the medium is non-relativistic and the contribution of the internal energy 
to the total energy density is then negligible as well, i.e., $\epsilon \approx 
\rho$. This approximation is consistent also with the assumption of negligible 
self-electromagnetic-field.

In order to find a solution for the pressure $p$, it is useful to rewrite 
equations (\ref{pressure_RN}) as equivalent equations for the density
\begin{eqnarray}
\label{density_RN}
\partial_r \rho&=&\frac{(\kappa \rho^{\Gamma}+\rho)\left(\partial_r\,\ln{|U_t|} - \frac{\Omega \partial_r \ell}{1-\Omega \ell}\right)-q\,U^t\partial_r A_t}{-\Gamma\kappa\rho^{\Gamma-1}},\nonumber\\
\partial_{\theta} \rho&=&\frac{(\kappa \rho^{\Gamma}+\rho)\left(\partial_{\theta}\,\ln{|U_t|} - \frac{\Omega \partial_{\theta} \ell}{1-\Omega \ell}\right)}{-\Gamma\kappa\rho^{\Gamma-1}},
\end{eqnarray}
 which can be solved more easily than the ones for the pressure. On the other 
hand, in the uncharged case, the pressure profiles are determined from relation 
(\ref{Boyer}) which, for a polytropic equation of state, gives the following 
relatively simple analytic formula
 \begin{eqnarray}
\label{pressure_RN0}
p=\left(\frac{{\rm e}^{\frac{\Gamma-1}{\Gamma}(W_{\rm in}-W)}-1}{\kappa^{\frac{1}{\Gamma}}}\right)^{\frac{\Gamma}{\Gamma-1}}.
\end{eqnarray}
 (Note that this formula is valid only in the region where $W_{\rm in}\geq W$).

In the next sections, we consider the commonly-used condition of constant 
specific angular momentum, $\ell={\rm const}$ 
\cite{Koz-Jar-Abr:1978:ASTRA:,Abr-Jar-Sik:1978:ASTRA:,Stu-Sla-Hle:2000:ASTRA:,Sla-Stu:2005:CLAQG:,Fon-Dai:2002:MNRAS:,Rez-Zan-Fon:2003:ASTRA:,Stu:2005:MODPLA:,Sla-Stu:2005:CLAQG:,Stu-Sla-Kov:2009:CLAQG:,Kuc-Sla-Stu:2011:CASTRP:}. 
Tori with $\ell=\mbox{const}$ are particularly simple mathematically and are 
generally representative of those with a more general angular momentum profile, 
although some care needs to be taken when considering perturbations (these 
configurations are only marginally stable with respect to convective 
instability \cite{Seg:1975}).
\section{Uncharged tori, $\boldsymbol{\ell={\rm const}}$}
 For a barotropic fluid, the isobaric surfaces coincide with the equipotential 
surfaces of the potential 
$W(r,\theta)$, which are given by the formula
 \begin{equation}                                                              
W(r,\theta)=\ln{|U_t|}=\ln{\frac{r\sqrt{\Delta}\sin{\theta}}{\sqrt{r^4\sin^2{\theta}-\ell^2\Delta}}}.
\end{equation} 
 One of the reality conditions here, $\Delta\geq0$, restricts the existence of 
equipotential surfaces to the stationary region of the spacetime. The other 
one, $r^4\sin^2{\theta}-\ell^2\Delta>0$, evaluated in the equatorial plane 
($\theta=\pi/2$), corresponds to the restriction that non-massless particles 
must move more slowly than photons 
 \begin{equation}                                                              
\ell^2 < \ell^2_{\rm ph}(r;Q^2)\equiv\frac{r^4}{\Delta},
\end{equation}
 where the function $\ell^2_{\rm ph}(r;Q^2)$ plays the role of the effective 
potential governing photon geodesic motion in the equatorial plane (see 
\cite{Stu-Hle:2000} where the more general case with a non-zero cosmological 
constant is discussed). For the purpose of classification, we only need to 
consider positive values of $\ell_{\rm ph}(r;Q^2)$, i.e., we define $\ell_{\rm 
ph}(r;Q^2)\equiv\frac{r^2}{\sqrt{\Delta}}$. The function $\ell_{\rm ph}(r;Q^2)$ 
has one local extremum (a minimum) outside of the outer black-hole horizon, 
$\ell_{\rm ph,c}(Q^2)$, located at $r_{\rm ph,c}=\frac{1}{2}(3+\sqrt{9-8Q^2})$, 
corresponding to the circular photon orbit in the equatorial plane.
\begin{figure}
\centering
\includegraphics[width=1\hsize]{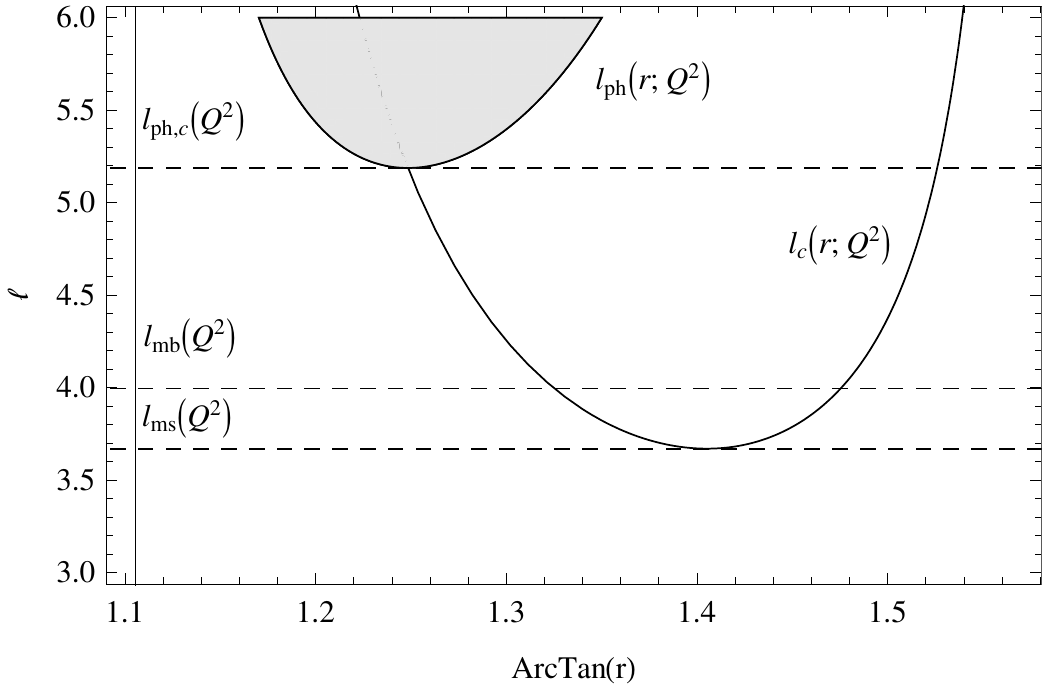}
 \caption{Behavior of the function $\ell_{\rm c}(r;Q^2)$ governing extrema of 
the potential $W(r,\theta)$ in the equatorial plane of the 
Reissner-Nordstr\"{o}m spacetime with $Q=0.1$. For a fixed value of the angular 
momentum $\ell$, we can determine the position of the potential maxima (smaller 
radius) and minima (larger radius). The potential $W(r,\theta)$ is not defined 
in the shaded region limited by the function $\ell_{\rm ph}(r;Q^2)$. In 
black-hole spacetimes ($Q^2\leq 1$), there is only the class of the mutual 
behavior of the functions $\ell_{\rm c}(r;Q^2)$ and $\ell_{\rm ph}(r;Q^2)$, 
shown in this figure. The horizontal dashed lines denote the values of 
$\ell_{\rm ms}(Q^2)\doteq 3.670$, $\ell_{\rm mb}(Q^2)\doteq 3.995$ and $\ell_{\rm 
ph,c}(Q^2)\doteq 5.187$ for $Q=0.1$, the value of $Q$ being considered here, and 
the vertical solid line shows the position of the outer black-hole horizon. 
From the discussion of the behavior of $W(r,\theta)$ (see Fig.~ \ref{Fig:2}), 
it follows that equilibrium toroidal configurations of a barotropic fluid with 
$\ell={\rm const}$ exist only for $\ell>\ell_{\rm ms}(Q^2)$.}
 \label{Fig:1}
\end{figure}

The character of the equipotential surfaces is well represented by the 
behavior of the potential $W(r,\theta)$ in the equatorial plane, i.e., by the 
function $W_{\pi/2}(r)\equiv W(r,\theta=\pi/2)$. Since the orbits with 
vanishing gradient of $W$, i.e. those satisfying the conditions $\partial_r 
W(r,\theta)=\partial_{\theta}W(r,\theta)=0$, correspond to loci with zero 
pressure gradients, the fluid has to follow geodesic motion there. The local 
extrema of $W$ are located only in the equatorial plane. Evaluating the 
necessary condition $\partial_{r}W_{\pi/2}(r)=0$, these extrema are given by 
the condition
 \begin{equation}
\label{extremaW}                                                               
\ell^2=\ell^2_{\rm c}(r;Q^2)\equiv\frac{r^4(r-Q^2)}{\Delta^2}.
\end{equation}
 Again, for the purpose of classification, we only need to consider positive 
values of $\ell_{\rm c}(r;Q^2)$. This function has one local minimum $\ell_{\rm 
ms}(Q^2)$ corresponding to the specific angular momentum of the marginally 
stable orbit. In addition to the limiting values $\ell_{\rm ph,c}(Q^2)$ and 
$\ell_{\rm ms}(Q^2)$, there is also another one, $\ell_{\rm mb}(Q^2)$, 
corresponding to the specific angular momentum of a particle moving along the 
marginally bound equatorial circular geodesic, determined by the conditions 
$\partial_r W_{\pi/2}(r)=0$ and $W_{\pi/2}(r)=0$.
\begin{figure*}
\centering
\includegraphics[width=1\hsize]{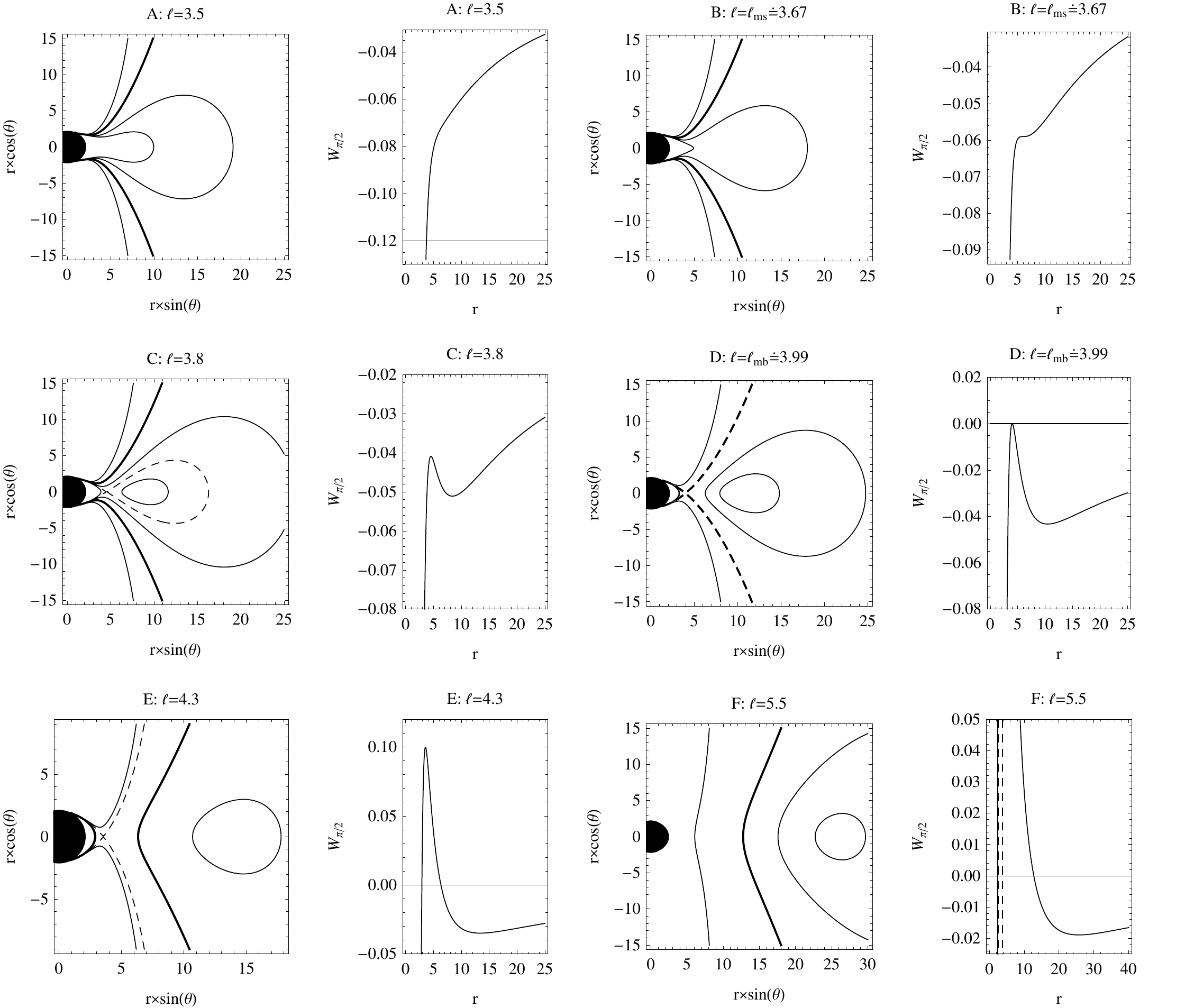}
 \caption{Typical behavior of the potential $W(r,\theta)$ shown in terms of 
poloidal sections through the equipotential surfaces and the equatorial profile 
$W_{\pi/2}(r)$ in a Reissner-Nordstr\"{o}m spacetime with parameter $Q=0.1$. 
Taking progressively increasing values of the specific angular momentum $\ell$, 
samples are shown of 4+2 qualitatively different types of behavior of the 
potential, differing in the properties of the critical (dashed) equipotential 
surface and the null (thick) equipotential surface.}
 \label{Fig:2}
\end{figure*}

Independently of the charge parameter $|Q|\leq1$ (we do not consider naked 
singularity spacetimes here), there is only one type of mutual behavior for 
$\ell_{\rm ph}(r;Q^2)$ and $\ell_{\rm c}(r;Q^2)$. We show this in 
Fig.~\ref{Fig:1}, drawn for $Q=0.1$ which we take as a standard value in the 
following. (Note that while this value for $Q$ is rather high from an 
astrophysical point of view \cite{Wal:1984}, it still gives only very small 
deviations of the space-time away from that of the Schwarzschild metric; taking 
a value this high is useful for clearly demonstrating the effects which we are 
wanting to investigate.)

Equilibrium toroidal configurations of a barotropic fluid with constant 
specific angular momentum exist only for $\ell>\ell_{\rm ms}(Q^2)$, which takes 
values ranging from $l_{\rm ms}(0)=3.674$, corresponding to the Schwarschild 
limit, to $l_{\rm ms}(1)=3.079$, corresponding to the extreme 
Reissner-Nordstr\"{o}m black hole limit. Moreover, an important feature of 
constant specific angular momentum tori is that those which can form a cusp are 
limited by $l<l_{\rm mb}(Q^2)$, which takes values ranging from $l_{\rm mb}(0) 
= 4$ to $l_{\rm mb}(1)= 3.330$. The limit $l_{\rm mb}(0)= 4$ for the 
Schwarzschild case was also found for self-gravitating tori \cite{Ste:2011}. 
The radii given by the condition $\ell=\ell_{\rm c}(r;Q^2)$ then correspond to 
motion of fluid elements along the unstable circular geodesic (smaller radius) 
and the stable one (larger radius). The stable circular geodesic represents the 
`center' of the torus (the potential $W(r,\theta)$ has a local minimum there 
while the pressure is maximal there). The unstable circular geodesic marks a 
critical point (cusp), where the potential $W(r,\theta)$ has a local maximum; 
the corresponding equipotential surface is self-crossing and is referred to as 
the `critical surface'. As well as this critical equipotential surface, there 
is also the characteristic null equipotential surface $W(r,\theta)=0$, which 
crosses the equatorial plane at infinity.

The behavior of the potential $W(r,\theta)$ can be summarized in the following 
way:\\
 \\ 
 For $\ell\in(0,\ell_{\rm ms})$, there are no extrema of the potential $W$,
and there are no closed equipotential surfaces and no critical equipotential 
surface. The null equipotential surface is open towards the black hole 
(Fig.~\ref{Fig:2}A).\\
 \\ 
 For $\ell=\ell_{\rm ms}$, there is one inflexion point of the potential $W$ 
in the equatorial plane at which the critical surface has its critical point, 
corresponding to a ring. The null equipotential surface is open towards the 
black hole (Fig.~\ref{Fig:2}B).\\
 \\
 For $\ell\in(\ell_{\rm ms},\ell_{\rm mb})$, there is a negative local 
maximum and a negative local minimum of the potential $W$ in the equatorial 
plane. In this case, closed equipotential surfaces exist which are bounded by 
the critical surface that self-crosses at the inner cusp. The null 
equipotential surfaces is open towards the black hole (Fig.~\ref{Fig:2}C).\\
 \\ 
 For $\ell=\ell_{\rm mb}$, there is a zero local maximum and a negative 
local minimum of the potential $W$ in the equatorial plane. The closed 
equipotential surfaces are bounded by the critical surface which coincides with 
the null equipotential surface (Fig.~\ref{Fig:2}D).\\
 \\
 For $\ell\in(\ell_{\rm mb},\ell_{\rm ph,c})$, there is a positive local 
maximum and a negative local minimum of the potential $W$ in the equatorial 
plane. The closed equipotential surfaces are bounded by the outer null 
equipotential surface. The critical surface is now open outwards away from the 
black hole, and self-crosses between the radii where the null surfaces cross 
the equatorial plane (Fig.~\ref{Fig:2}E).\\
 \\
 For $\ell=\ell_{\rm ph,c}$, the potential $W$ diverges at $r_{\rm ph,c}$ and 
the local maximum no longer exists. The negative local minimum of the potential 
$W$ is still present. The closed equipotential surfaces are bounded by the 
outer null equipotential surface. The critical surface is no longer present.\\
 \\
 For $\ell>\ell_{\rm ph,c}$, the only extremum of the potential $W$ is the 
negative minimum. The closed equipotential surfaces are bounded by the outer 
null equipotential surface. There is no longer any critical surface, but there 
is a forbidden region for fluid elements with prescribed specific angular 
momentum, delimited by the radii satisfying the relation $\ell=\ell_{\rm 
ph}(r;Q^2)$ (Fig.~\ref{Fig:2}F).\\
 \\
 The behavior of the potential $W(r,\theta)$ is qualitatively the same as in 
the pure Schwarzschild case \cite{Abr-Jar-Sik:1978:ASTRA:}. The additional 
charge of the black hole $Q$ here influences the values of $\ell_{\rm 
ms}(Q^2)$, $\ell_{\rm mb}(Q^2)$ and $\ell_{\rm ph,c}(Q^2)$.
 
Profiles of the pressure and mass-density can now be determined from relations 
(\ref{pressure_RN0}) and (\ref{EOS}). For doing this, it is necessary to choose 
relevant values for the parameters of the polytropic equation of state 
(\ref{EOS}), for the specific angular momentum and for the location of the 
inner edge of the torus. Here we present two examples, for tori 
with $\Gamma=2$, $\kappa=10^{12}$ and $\ell=3.8$ (Fig.~\ref{Fig:3}): 
\begin{figure*}
\centering
\includegraphics[width=1\hsize]{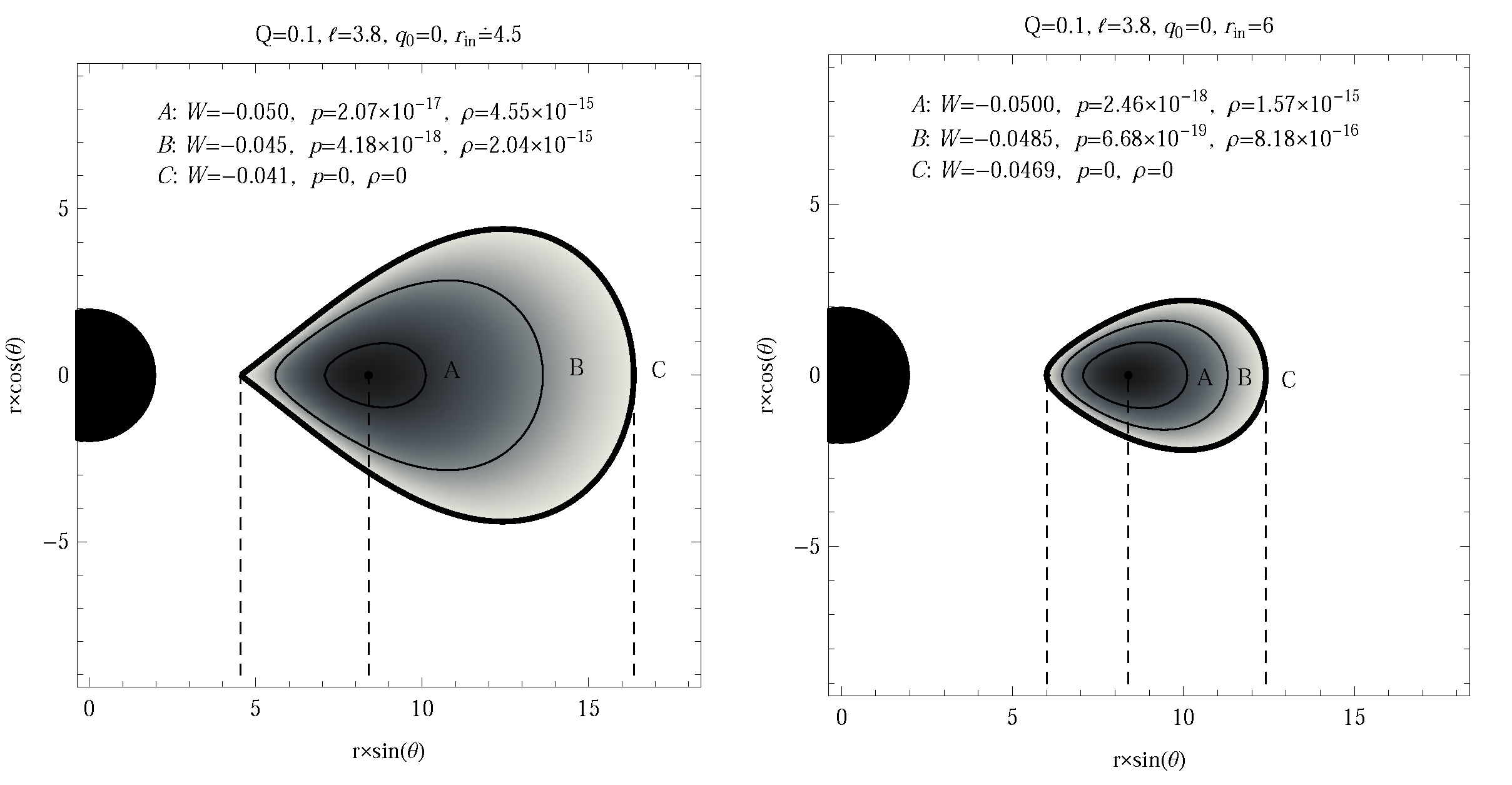}
 \caption{Profiles of the potential $W(r,\theta)$, pressure $p(r,\theta)$ and 
rest-mass density $\rho(r,\theta)$ of uncharged tori, shown in terms of 
poloidal sections through the equipotential, isobaric and iso-density surfaces, 
respectively, and their equatorial behavior $W_{\pi/2}(r)$, $p_{\pi/2}(r)$ and 
$\rho_{\pi/2}(r)$, for a Reissner-Nordstr\"{o}m spacetime with $Q=0.1$. Two 
examples are shown with different values for the radius at the inner edge of 
the torus: $r_{\rm in}=r_{\rm cusp}\doteq4.544$ in the upper figures and 
$r_{\rm in}=6$ in the lower ones. The centers (pressure maxima) of both 
tori are located at $r_{\rm cent}\doteq 8.388$. The shapes of the 
equipotential, isobaric and iso-density surfaces coincide, with the values of 
the different quantities on them being related by equation (\ref{pressure_RN0}) 
and the equation of state (\ref{EOS}). We show poloidal sections through three 
such surfaces (A, B and C). The thick curve (surface C) marks the zero-pressure 
(and zero-density) surface which bounds the torus. The shaded regions above the 
profiles of $W_{\pi/2}(r)$ indicate the physically relevant parts of the 
profiles, delimited by the inner and outer edges of the tori. The specific 
angular momentum $\ell=3.8$ for both tori, and the polytropic parameters are 
$\Gamma=2$ and $\kappa=10^{12}$ in each case.}
 \label{Fig:3}
\end{figure*}

The first is the marginally bounded torus (the most extended closed 
torus), with its inner edge being at the cusp, $r_{\rm in}=r_{\rm cusp}$. This 
is the most interesting case, since it can be used as a model for the inner 
parts of a thick accretion disk from which matter can flow in a standard way 
onto the black hole. To obtain the position of the cusp, we solve equation 
(\ref{extremaW}), which yields two real roots above the event horizon; $r_{\rm 
I}\doteq 4.544$ and $r_{\rm II}\doteq 8.388$. The first root corresponds to the 
position of the unstable circular geodesic, i.e. to the cusp, while the second 
one corresponds to the stable geodesic, i.e to the pressure maximum: the 
`center' of the torus.

Our second example is a torus with its inner edge located at $r_{\rm in}>r_{\rm 
cusp}$. Here we simply chose the position of the inner edge to be at $r_{\rm 
in}=6$. Note that the location of the inner edge must be chosen so as to be in 
between the cusp ($r_{\rm I}$) and the center of the torus ($r_{\rm II}$), the 
positions of both of which are determined from the potential $W$ independently 
of the choice of $r_{\rm in}$.

\section{Charged tori, $\boldsymbol{\ell={\rm const}}$}
 In order to obtain the pressure or density profiles from equations 
(\ref{pressure_RN}) or (\ref{density_RN}), it is necessary to determine the 
charge density distribution $q(r,\theta)$ in the torus. This must satisfy the 
integrability condition (\ref{condition}). By expressing $q(r,\theta)$ in the 
form
 \begin{eqnarray}
\label{chargedistribution}
q(r,\theta)=q_0\rho(r,\theta)k(r,\theta),
\end{eqnarray}
 where $q_0$ is a constant and $k(r,\theta)$ is a correction function, and 
using $\Gamma=2$, we can rewrite equation (\ref{density_RN}) in the form
 \begin{eqnarray}
\label{density_RN_2}
\partial_r \rho&=&-\frac{1}{2\kappa}\bigg((\kappa \rho+1)\partial_r\,\ln{|U_t|}-q_0k\,U^t\partial_r A_t\bigg),\nonumber\\
\partial_{\theta} \rho&=&-\frac{1}{2\kappa}\bigg((\kappa \rho+1)\partial_{\theta}\,\ln{|U_t|}\bigg).
\end{eqnarray}
Note that, as we express by relation (\ref{chargedistribution}), it is 
feasible to take the charge density distribution as being directly proportional 
to the rest-mass density distribution and to a correction function which 
represents variations in the charge per unit mass $q_0 k(r,\theta)$ through the 
torus, as we discuss in section VI A. From the mathematical point of view, the 
correction function plays the role of an `integration factor', which must be 
chosen so that equations (\ref{density_RN_2}) are integrable.

Now, due to the integrability condition (\ref{condition}), the correction 
function $k(r,\theta)$ has to satisfy the relation
 \begin{eqnarray}
2\sin{\theta}(\ell^2\Delta-r^4\sin^2{\theta})\partial_{\theta} k+3k\ell^2\Delta \cos{\theta}=0,
\end{eqnarray}
which can be solved to give
 \begin{eqnarray}
\label{correction}
k(r,\theta)=\frac{\gamma}{\sin^{3/2}{\theta}}\left(\frac{\ell^2\Delta-r^4\sin^2{\theta}}{\ell^2\Delta-r^4}\right)^{3/4},
\end{eqnarray} 
 where $\gamma(r)$ is a function representing a constant of integration over 
$\theta$ for a given value of $k$ in the equatorial plane, 
$k(r,\theta=\pi/2)\equiv\gamma(r)$. For the purposes of this paper, it is 
convenient to choose $\gamma(r)=1$. The charge density distribution function 
with the correction function in the form (\ref{correction}) ensures the 
integrability of equations (\ref{density_RN_2}), and thus the existence of a 
unique solution for $p(r,\theta)$ and $\rho(r,\theta)$.

Integrating the second of equations (\ref{density_RN_2}) over the latitude, we 
obtain the following expression for the rest-mass density:
 \begin{eqnarray}
\label{rho1}
\rho(r,\theta)=\frac{2^{1/4}\kappa\,C\left(r^4\sin^2{\theta}-\ell^2\Delta\right)^{1/4}-\sqrt{\sin{\theta}}}{\kappa \sqrt{\sin{\theta}}}.
\end{eqnarray}
 The unknown function $C(r)$, which depends only on the radial coordinate $r$, 
stands as a constant of this integration. Its value can be determined by 
substituting the density formula (\ref{rho1}) into the first density equation 
(\ref{density_RN_2}) and assuming the charge-density distribution according to 
relations (\ref{chargedistribution}) and (\ref{correction}). This leads to the 
ordinary differential equation
 \begin{eqnarray}
\label{rho2}
2r\Delta\partial_r C+(2r^2-3r+Q^2)C=\frac{r^2\sqrt{\Delta}Qq_0}{2^{1/4}\kappa (r^4-\ell^2\Delta)^{3/4}}.\nonumber\\
\end{eqnarray} 
 Unfortunately, there is no analytic solution for equation (\ref{rho2}), and 
so $C(r)$ must be determined numerically. Since the torus is delimited by the 
zero-pressure (and zero-density) surface, we can find the necessary initial 
condition from the fact that $\rho(r_{\rm in})=0$. From the relation 
(\ref{rho1}) we obtain
 \begin{eqnarray}
\label{initial}
C(r_{\rm in})=\frac{1}{2^{1/4}\kappa[r_{\rm in}^4-\ell^2(r_{\rm in}^2-2r_{\rm in}+Q^2)]^{1/4}}.
\end{eqnarray}
\begin{figure*}
\centering
\includegraphics[width=0.85\hsize]{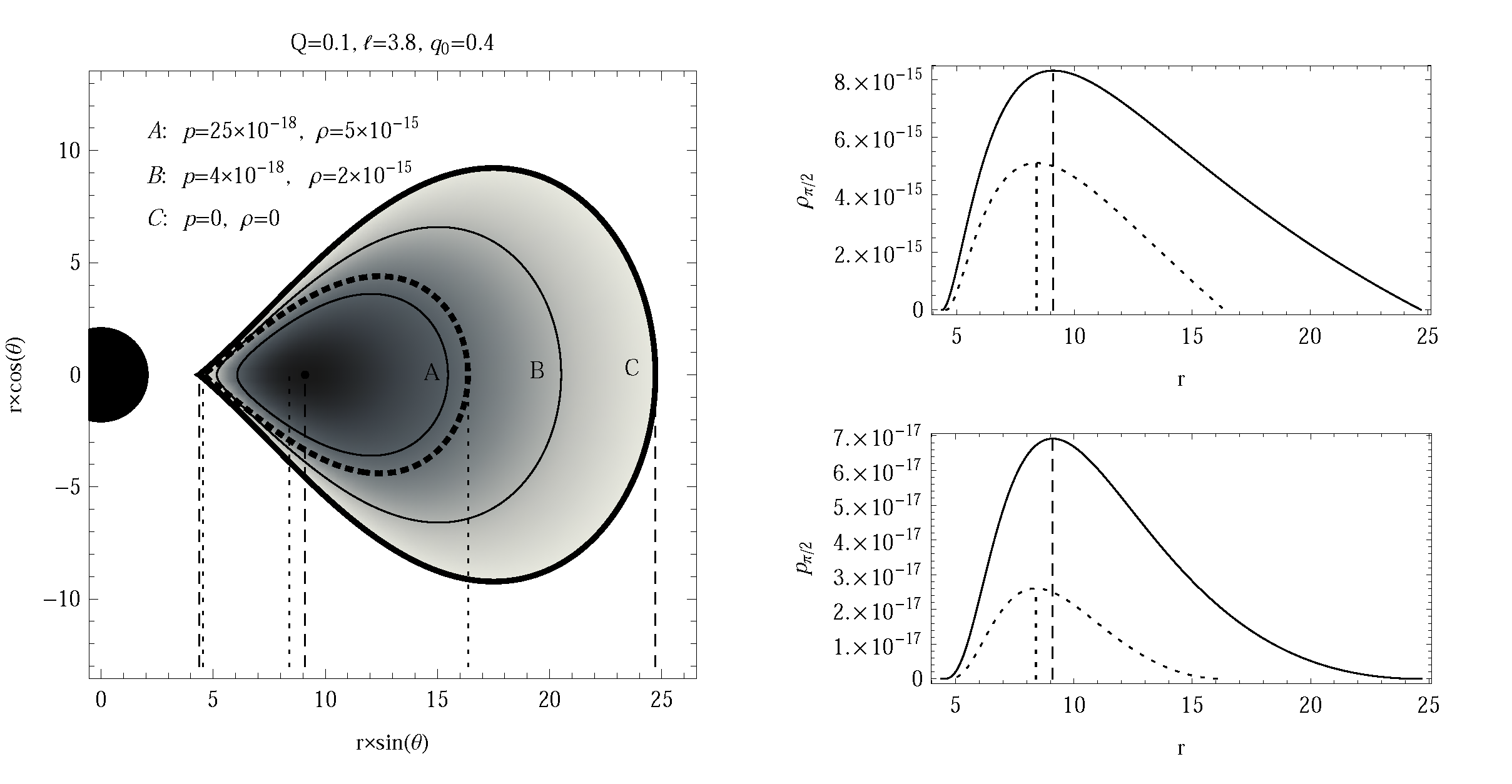}
 \caption{Profiles of the pressure $p(r,\theta)$ and rest-mass density 
$\rho(r,\theta)$ for a positively charged torus ($q_0=0.4$, $\ell=3.8$) in a 
Reissner-Nordstr\"{o}m spacetime with $Q=0.1$, shown in terms of poloidal 
sections through the isobaric and iso-density surfaces (left), and in terms of 
their equatorial profiles $p_{\pi/2}(r)$ and $\rho_{\pi/2}(r)$ (right). The 
torus terminates at $r_{\rm in}=r_{\rm cusp}\doteq 4.378$ and $r_{\rm out}\doteq 
24.72$, and its center is located at $r_{\rm cent}\doteq 9.098$. The dashed curves 
represent the zero-pressure surface for the equivalent uncharged case (left 
panel) and the profiles of density and pressure for the uncharged case (right 
panels). The isobaric and iso-density surfaces coincide (A, B and C), with the 
values of pressure and density on them being related by the equation of state.
}
 \label{Fig:4}
\end{figure*}
\begin{figure*}
\centering
\includegraphics[width=0.85\hsize]{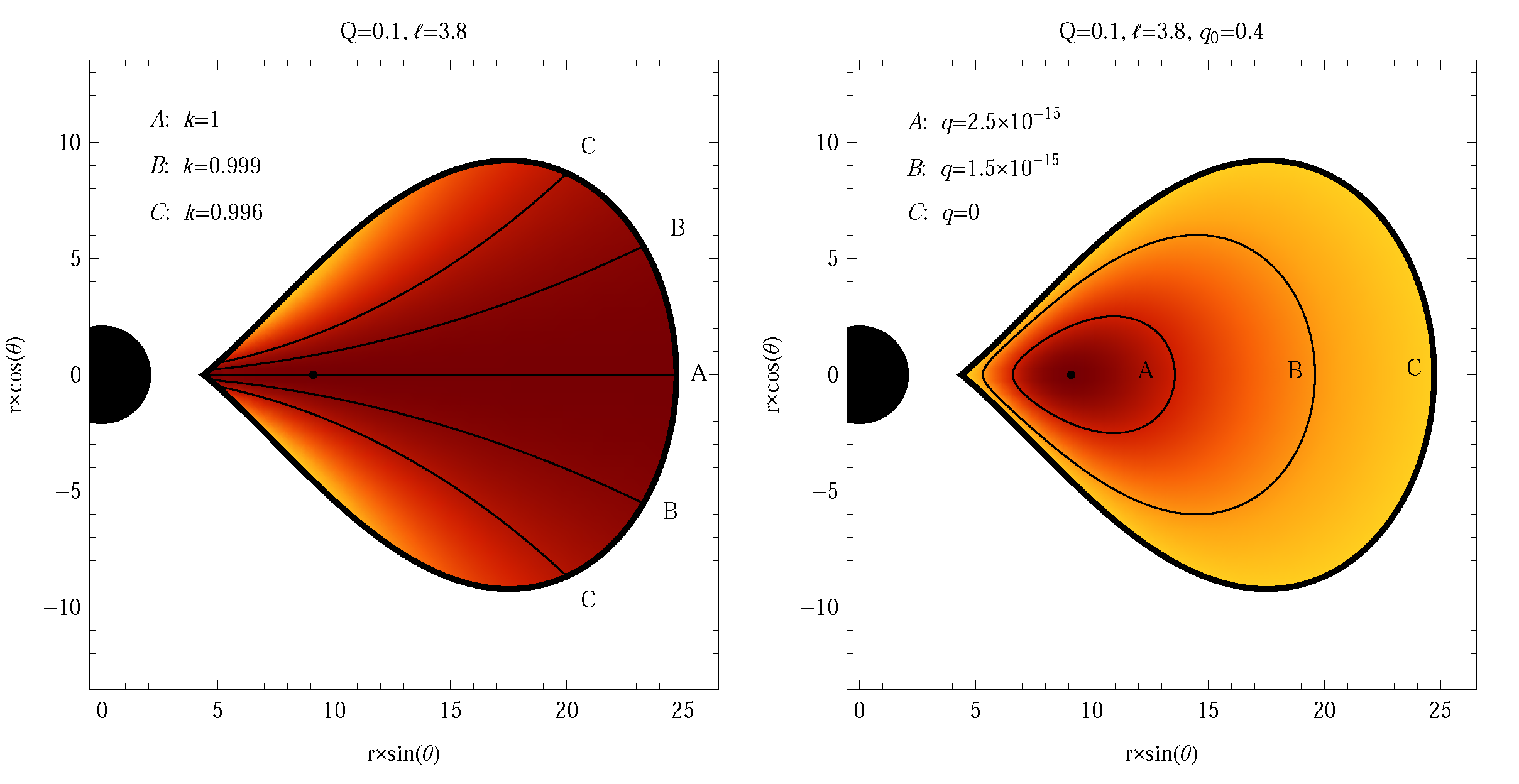}
 \caption{Poloidal sections through the iso-contours for the correction
function $k(r,\theta)$ (left) and the charge density $q(r,\theta)$ (right) for
the same positively charged torus as in Fig.~\ref{Fig:4}.}
 \label{Fig:5}
\end{figure*}
\begin{figure*}
\centering
\includegraphics[width=0.85\hsize]{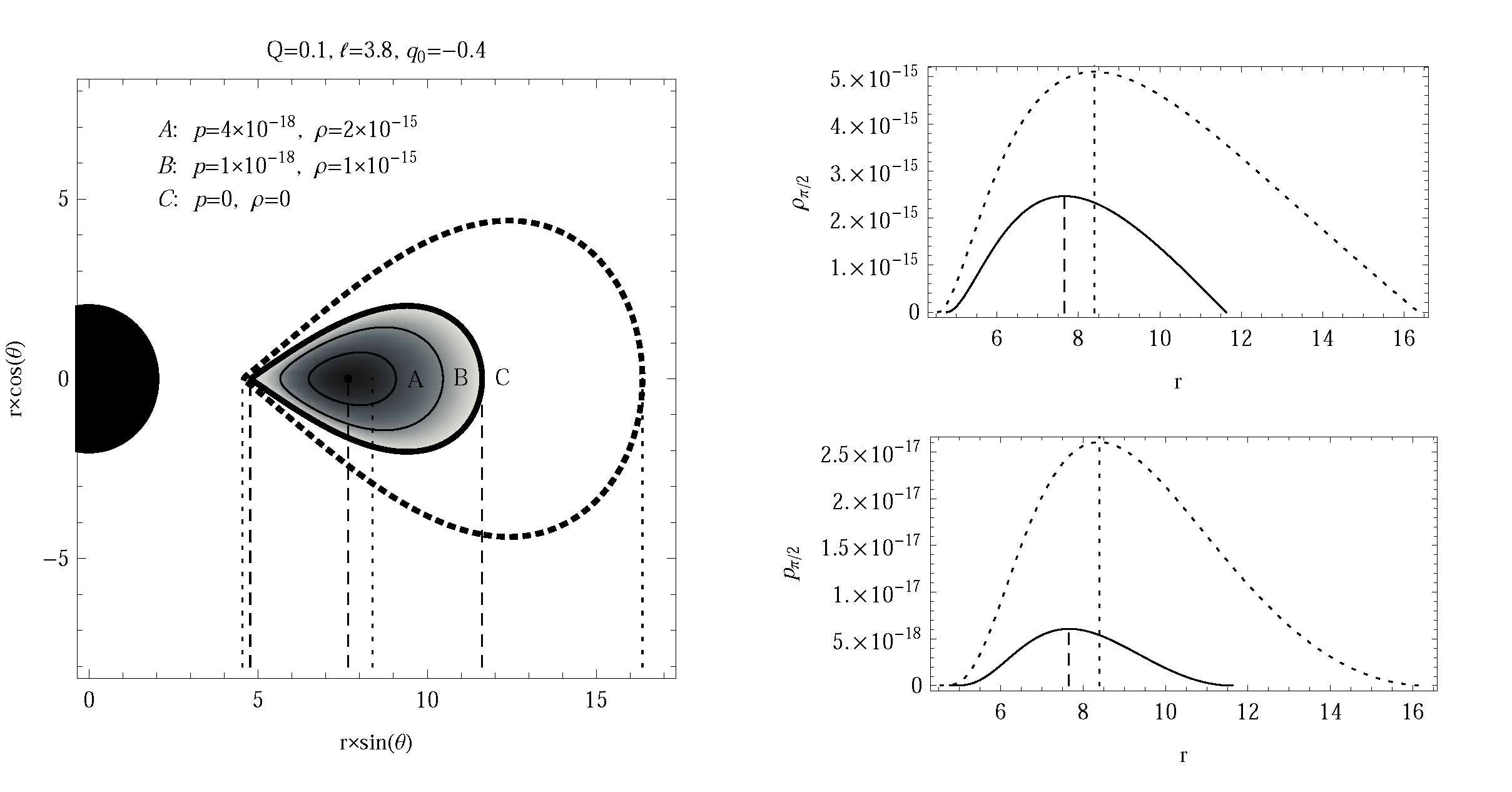}
 \caption{Profiles of the pressure $p(r,\theta)$ and rest-mass density 
$\rho(r,\theta)$ for a negatively charged torus ($q_0=-0.4$, $\ell=3.8$) in a 
Reissner-Nordstr\"{o}m spacetime with $Q=0.1$, shown in terms of poloidal 
sections through the isobaric and iso-density surfaces (left), and in terms of 
their equatorial profiles $p_{\pi/2}(r)$ and $\rho_{\pi/2}(r)$ (right). The 
torus terminates at $r_{\rm in}=r_{\rm cusp}\doteq 4.767$ and $r_{\rm out}\doteq 
11.62$, and its center is located at $r_{\rm cent}\doteq 7.660$. The dashed 
curves represent the zero-pressure surface for the equivalent uncharged case 
(left panel) and the profiles of density and pressure for the uncharged case 
(right panels). The isobaric and iso-density surfaces coincide (A, B and C), 
with the values of pressure and density on them being related by the equation 
of state.}
 \label{Fig:6}
\end{figure*}
\begin{figure*}
\centering
\includegraphics[width=0.85\hsize]{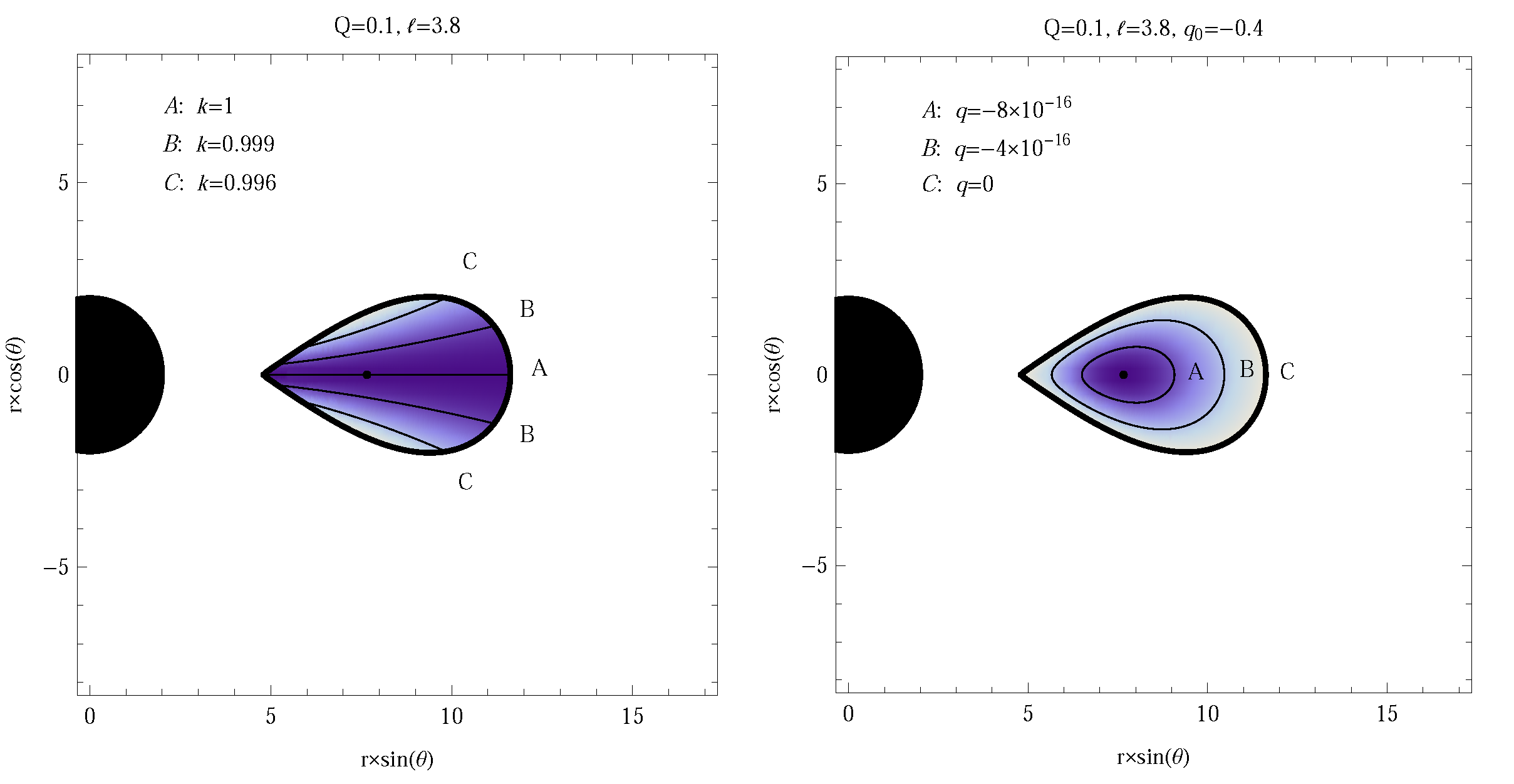}
 \caption{Poloidal sections through the iso-contours for the correction 
function $k(r,\theta)$ (left) and the charge density $q(r,\theta)$ (right) for 
the same negatively charged torus as in Fig.~\ref{Fig:6}.}
 \label{Fig:7}
\end{figure*}

As mentioned earlier, the most interesting configuration is the one 
delimited by the self-crossing zero isobaric (and iso-density) surface, i.e., 
the marginally bounded torus, which has its inner edge in the equatorial plane 
coincident with the cusp ($r_{\rm in} = r_{\rm cusp}$). We will be 
concentrating on this type of torus from now on. The values of the specific 
angular momentum $\ell$ and the charges $Q$ and $q_0$ then completely determine 
the shape of the torus and the positions of its center and of its inner and 
outer edges. At the cusp, the pressure and density vanish and have a saddle 
point (a minimum in the \mbox{$r$-direction} and a maximum in the 
\mbox{$\theta$-direction}), in contrast with the center of the torus, where 
there is a local maximum. (In general, the density at the inner edge of the 
torus must be zero and when the inner edge is also a cusp, the density profile 
has extrema there.) The location of the inner edge $r_{\rm in}$, which needs to 
be known in order to evaluate the condition (\ref{initial}), can then be 
obtained from the first of equations (\ref{density_RN_2}) by setting 
$\partial_r \rho\big|_{r=r_{\rm in}}=0$, $\rho(r_{\rm in})=0$, and 
$\theta=\pi/2$. This gives the following implicit expression for $r_{\rm in}$:
 \begin{eqnarray}
\label{q0l}
q_0 Q r_{\rm in}^2\sqrt{\Delta(r_{\rm in}^4-\ell^2\Delta)}+(Q^2-r_{\rm in})r_{\rm in}^4+\ell^2\Delta^2=0. 
\end{eqnarray}  
 Note that when $q_0=0$, we get the relation $(Q^2-r_{\rm in})r_{\rm 
in}^4+\ell^2\Delta^2=0$, in agreement with the formula (\ref{extremaW}) derived 
for the case of the uncharged torus.

To clearly illustrate how the charge on the torus affects its equilibrium 
structure, we constructed marginally-bounded tori with the same matter 
parameters and specific angular momentum as in the uncharged case, i.e., we 
took $\kappa=10^{12}$ and $\ell=3.8$, and we also considered the same 
charge of the central black hole $Q=0.1$. Our sample tori are characterized by 
the parameters $q_0=0.4$ (positively charged) and $q_0=-0.4$ (negatively 
charged), which, as we discuss in section VI A, correspond to the specific 
charges of the moving matter in the equatorial plane, due to our choice 
$\gamma(r)=1$.

For $q_0=0.4$, equation (\ref{q0l}) yields two real roots above the event 
horizon; $r_{\rm I}\doteq 4.378$ and $r_{\rm II}\doteq 9.104$, where both of 
the roots correspond to circular orbits of charged test particles with the 
given specific charge $0.4$. Choosing $r_{\rm in}=r_{\rm I}$, the numerical integration gives a
regular thick charged torus. On the other 
hand, choosing $r_{\rm in}=r_{\rm II}$, we get a degenerate torus (an 
infinitesimally thin ring) located just at $r=r_{\rm II}$.

For $q_0=-0.4$, equation (\ref{q0l}) again yields two real roots above the 
event horizon; $r_{\rm I}\doteq 4.767$ and $r_{\rm II}\doteq 7.658$, 
corresponding to circular orbits of charged test particles with the given 
specific charge $-0.4$. Again, the choice $r_{\rm in}=r_{\rm I}$ leads to the 
regular thick charged torus, while $r_{\rm in}=r_{\rm II}$ gives the degenerate 
torus.

In principle, one could also construct charged tori with $r_{\rm I}<r_{\rm 
in}<r_{\rm II}$, as for the uncharged tori. However, this introduces further 
complications, and we focus here only on the more interesting critical (cusp) 
configurations.

In Figs \ref{Fig:4} and \ref{Fig:6}, we show the profiles of rest-mass density 
and pressure. The related charge density distributions $q(r,\theta)$ and 
correction functions $k(r,\theta)$ are shown in Figs \ref{Fig:5} and 
\ref{Fig:7}. As can be seen from these, for the same $Q$ and $\ell$, the 
positively charged tori are more extended than the uncharged ones while the 
negatively charged tori are less extended. From the right-hand panels of Figs 4 
and 6, it can be seen that the densities in the more extended tori are larger 
than those in the less extended ones and so the masses of the more extended 
tori are clearly also larger. We come back to this in more detail in 
section VI B.
\section{Discussion}
\subsection{Correction function and specific charge}
 The dimensionless correction function $k(r,\theta)$ has been introduced in 
equation (\ref{chargedistribution}) for convenience in calculating the torus 
configuration. It represents the variation with position of the specific 
electric charge (charge per unit mass), $\bar{q}=q_0 k(r,\theta)$. As shown in 
the previous section, setting $k=1$ so that the charge per unit mass is the 
same everywhere, does not give an equilibrium configuration: $k$ must be 
allowed to vary so as to satisfy the integrability condition (\ref{condition}) 
and this requires the behavior (\ref{correction}). In the equatorial plane 
$k(r,\theta=\pi/2)=\gamma(r)$, and the choice $\gamma(r)=1$ as a boundary 
condition is convenient for our present simplified model. From Figs 
(\ref{Fig:5}) and (\ref{Fig:7}), it can be seen that the required variations in 
$k(r,\theta)$ away from this are actually very small (with the maximum being on 
the equatorial plane).

Our choice of $\gamma(r)={\rm const}$ (with the constant normalised to 1) 
corresponds to a case where the maximum of the charge density $q(r,\theta)$ is 
located just at the center of the torus, where there is the maximum of the 
density $\rho(r,\theta)$, as can be seen from relation (28). Other choices of 
$\gamma(r)$ could describe more physically relevant situations, but with the 
maximum of $q(r,\theta)$ not necessarily being located at the center of the 
torus. For instance, by choosing $\gamma(r)=1/r$ the specific net charge of the 
fluid in the torus grows in the equatorial plane from the outer edge to the 
inner edge, where it is maximal. Moreover, the maximum of $q(r,\theta)$ is 
shifted from the center of the torus. Of course, at the inner edge, the net 
charge density $q(r,\theta)$ goes to zero together with the matter density 
$\rho(r,\theta)$.

The values $q_0= \pm \, 0.4$ and $\gamma=1$, used for our representative cases, 
give a specific charge $\bar{q}$ in the torus around $10^{18}$ times smaller in 
magnitude than that for a proton, and so the medium can be thought of as having 
one particle in $10^{18}$ with a net charge while the rest are neutral. We note 
that if we decrease $Q$ below our standard value of $0.1$ and simultaneously 
increase $q_0$ in such a way that the product $Q\bar{q}$ remains unchanged, 
then we get essentially identical results. This is because the deviation of the 
space-time geometry away from Schwarzschild is extremely small for these values 
of $Q$, and so the relevant effect of the charge is almost entirely 
electromagnetic (depending on $Q\bar{q}$) rather than having a significant 
gravitational contribution (depending just on $Q$).

\subsection{Total electric charge and mass of the torus}
 In our model, we neglect the effects of the electromagnetic field generated by 
the torus, which is acceptable when this self-field is much weaker than the 
external electromagnetic field associated with the central compact object. The 
total charge of the torus is given by
 \begin{eqnarray}
\mathcal{Q}&=&\int_{V}q\sqrt{-g}\,\rm{d}{\it r}\,\rm{d}\theta\,\rm{d}\phi\nonumber\\
&=&4\pi q_0\int_{r_{\rm in}}^{r_{\rm out}}\int_{\theta_{\rho_0}}^{\frac{\pi}{2}}k\rho\sqrt{-g}\,\rm{d}\theta\,{\rm d}{\it r},
\end{eqnarray}
where 
\begin{eqnarray}
\theta_{\rho_0}=\arcsin\left(\frac{\sqrt{2\Delta}\kappa^2\ell C^2}{\sqrt{2\kappa^4r^4C^4-1}}\right)
\end{eqnarray}
 is the function determining the upper boundary of the poloidal projection of 
the zero isodensity surface, $g=-r^4\sin^2\theta$ is the determinant of the 
metric tensor, and $r_{\rm out}$ is the radial position of the outer edge of 
the torus in the equatorial plane. For the representative tori which we are 
considering, with $\Gamma=2$, $\kappa=10^{12}$, $Q=0.1$ and $\ell=3.8$, we 
obtain the total charge on the torus as being $\mathcal{Q}\doteq 2.49\times 
10^{-11}$ for $q_0=0.4$ (positively charged) and $\mathcal{Q}\doteq 
-4.50\times10^{-13}$ for $q_0=-0.4$ (negatively charged). Such small values, 
in comparison with the charge of the black hole, are consistent with our neglect 
of the electromagnetic field generated by the torus.

The total rest-mass of the torus is given by 
\begin{eqnarray}
\mathcal{M}&=&\int_{V}\rho\sqrt{-g}\,\rm{d}{\it r}\,\rm{d}\theta\,\rm{d}\phi\nonumber\\
&=&4\pi \int_{r_{\rm in}}^{r_{\rm out}}\int_{\theta_{\rho_0}}^{\frac{\pi}{2}}\rho\sqrt{-g}\,\rm{d}\theta\,{\rm d}{\it r}.
\end{eqnarray}
 For the three representative tori which we are considering, we obtain the 
total rest-mass as being $\mathcal{M}\doteq 6.24\times 10^{-11}$ (positively 
charged), $\mathcal{M}\doteq 1.13\times10^{-12}$ (negatively charged) and 
$\mathcal{M}\doteq 9.64\times 10^{-12}$ (uncharged). The ratio 
$\mathcal{Q}/\mathcal{M}$ is then $\doteq \pm 0.4$ in the positively and 
negatively charged cases, respectively, as clearly follows from the fact that 
the specific charge $\bar{q}=q_0 k(r,\theta)=\pm 0.4 k(r,\theta)\doteq \pm 
0.4$, since $k(r,\theta)\approx 1$ throughout our tori.

\subsection{Parameters of the equation of state}
 We set $\Gamma=2$ for illustration purposes, since this choice simplifies 
the integration of the density equations (\ref{density_RN}). This value of 
$\Gamma$ is inconveniently high for possible related astrophysical 
applications, but we stress that our model is an extremely simplified one, 
purposely intended for investigating the behavior of a test case under extreme 
conditions (we are also taking a rather large value for the black-hole charge 
$Q$, only one sign of charge for particles in the torus and zero conductivity 
there). For such a test case, it would not be appropriate to introduce 
additional complications here in order to bring just one aspect of the model 
(the value of $\Gamma$) closer to astrophysical applications.

In general, electrostatic corrections should also be included in the equation 
of state, especially at higher matter densities. However, this is not trivial 
to do (see, e.g. \cite{Avi:2006}, for the electrostatic correction in the case 
of dusty plasmas). We have used a very high value of $\kappa$, which enables us 
to neglect electrostatic corrections without inconsistency, because this 
enables us to demonstrate our approach more clearly.

\subsection{Distribution of specific angular momentum}
 We have considered perfect-fluid tori with a prescribed specific angular 
momentum $\ell=-U_{\phi}/U_t$, which we set to be constant through the torus. 
In an uncharged case, $L=U_{\phi}$ and $E=-U_t$ would be constants of motion 
connected with the assumed axial symmetry and stationarity of the spacetime. 
For a charged torus, the constants of motion are the generalized quantities 
$\tilde{L}=U_{\phi}+\bar{q}A_{\phi}$ and $\tilde{E}=-U_{t}-\bar{q}A_{t}$.

Note that the condition $\ell={\rm const}$ is imposed for simplicity of the 
calculations; it is not essential for the method and can be relaxed. The 
Rayleigh criterion for linear stability against radial convection requires 
$\ell$ to be a non-decreasing function of the distance from the axis of 
rotation, and so $\ell={\rm const}$ uncharged tori are just on the stability 
limit. For charged tori, the stability condition needs to be formulated in 
terms of a generalized quantity $\tilde{\ell}=\tilde{L}/\tilde{E}$ 
\cite{Vok-Kar:1991a,Abr-etal:1993}. In the Reissner-Nordstr\"{o}m electric 
field, the only non-zero component of the vector potential is $A_t$ and one has
 \begin{eqnarray}
\tilde{\ell}=\frac{\tilde{L}}{\tilde{E}}=\frac{L}{E-q_0kA_t},
\end{eqnarray}
 which reduces to $\ell$ for $q_0=0$. Stability depends heavily on the specific 
charge distribution (i.e. on the behavior of the correction function 
$k(r,\theta)$) and on the signs of the charges of the torus and the black hole.

\subsection{Approximation of negligible conductivity}
 The assumption of high electrical conductivity of the medium is appropriate 
for many astrophysical plasmas with a high degree of ionization, and the ideal 
MHD framework can then be employed. Under the conditions of high conductivity 
and vanishing inertial effects of the plasma particles, the local electric 
field quickly becomes neutralized by rearranging the plasma flows (giving the 
conditions for the force-free approximation). A quasi-neutral medium then 
arises in which the volume density of net electric charge is negligible.

However, there is an ongoing debate about the conditions that may lead to the 
presence of non-vanishing net charges, with an important role being played by 
electric forces acting parallel to the magnetic field lines in the local 
co-moving frame. For example, a large-scale magnetic field may cause spatial 
separation of electric charges of different signs and their gradual 
accumulation in different parts of the system. Pulsar magnetospheres provide an 
example of such systems, with the charge separation being caused by the 
dipole-type magnetic field of the neutron star \cite{Neu:1993,Pet-etal:2002}. 
Black holes embedded in ordered magnetic fields of external origin can also act 
in a similar way but then, for a low-density medium, the hydrodynamical 
description needs to be modified in order to describe the conditions of a 
collisionless plasma (since the particle mean-free paths are then comparable 
with the characteristic length-scale of the system, given by the gravitational 
radius of the central black hole).

One can imagine also another relevant scenario: a neutral fluid containing a 
few free charges, such as the case of dusty plasmas. In general, when charges 
feel an external electromagnetic field, they move generating a current, but if 
the fluid is dense enough and is highly collisional, the charges are less able 
to move and the conductivity becomes almost zero (see \cite{Rei:1965} for a 
discussion). Such a picture is actually compatible with our model since, as 
mentioned earlier, only a very small fraction of particles with net charge is 
required in order to give the parameter values used in our representative 
examples.

\subsection{Zero conductivity and consistency of the model}
 The various limiting situations which we have been mentioning (hydrodynamical 
versus collisionless plasma; infinite conductivity versus zero conductivity; 
self-gravitating matter versus test particles and fluids), are relevant under 
quite different circumstances and obviously require different approaches. The 
approach which we have adopted in the present paper allows us to capture the 
behavior of an idealized but non-trivial system where the fluid motion is 
governed by the combined action of a global (large-scale) electromagnetic 
field, the gravitational field of the central body, and pressure gradients 
operating within the fluid together with a non-zero electric charge 
distribution.

The basic assumptions of our model are: 1) the fluid is a single-component test 
fluid (we ignore its self-gravity and its own electromagnetic field); 2) the 
fluid flow is stationary, with the 4-velocity having only time and azimuthal 
components. If the conductivity $\sigma$ were non-zero, the second term on the 
right-hand side of the Ohm's law equation (\ref{Ohm}), which is proportional to 
$\sigma$, would give rise to a radial electric current unless there were a 
significant self-field (contradicting the first basic assumption). Since our 
fluid is taken to be a single-species one, having a radial electric current 
would imply the existence of a radial mass current (contradicting the second 
basic assumption). Having $\sigma=0$ is therefore necessary for 
self-consistency.
\section{Conclusions}
 In this paper we have presented a model for a simple test case of an 
electrically-charged perfect-fluid torus rotating in strong gravitational and 
electromagnetic fields produced by a central compact object. Distributions of 
either specific angular momentum or charge density through the torus first need 
to be specified (we chose to specify the specific angular momentum 
distribution) and then pressure and density profiles can be calculated. An 
equation of state must be provided in order to close the set of equations. We 
have investigated the limiting case opposite to that of ideal 
magnetohydrodynamics, considering a non-conductive (dielectric) perfect fluid 
rather than infinite conductivity as in ideal magnetohydrodynamics. Our case 
can describe a fluid in which bulk hydrodynamic motion predominates over 
electromagnetic effects and the fluid has almost infinite electric resistivity. 
We are not here including the self-gravitational and self-electromagnetic 
fields of the matter in the torus, and so the treatment applies for low-mass, 
slightly-charged tori.

For illustrating the model, we constructed both positively and negatively 
charged barotropic tori, with a polytropic equation of state and constant 
specific angular momentum distribution, encircling a positively-charged 
Reissner-Nordstr\"{o}m black hole. We compared the resulting pressure and 
density profiles with an equivalent uncharged case, and also calculated the 
shapes of the tori. Taking the polytropic index $\Gamma=2$, allows for the 
`pressure' equation to be integrated in a relatively simple way so as to give 
semi-analytic results. The large value for the polytropic constant $\kappa = 
10^{12}$ leads to tori where the Coulomb interaction between the charged particles of the fluid can be neglected in comparison with the 
standard pressure due to the matter. The constructed tori are only slightly electrically charged in 
comparison with the charge of the black hole, thus generating a relatively weak 
electromagnetic field which can safely be neglected. It is striking that even 
with a very small value for the charge-to-mass ratio in the torus, significant 
effects are nevertheless seen. However, it is necessary to stress out that the constructed tori carry the specific charge $\mathcal{Q}/\mathcal{M}\doteq\pm 0.4$ and orbit the black hole with $Q/M=0.1$; since for any astrophysical body it would be practically
impossible to maintain the specific charge $Q/M > 10^{-18}$ \cite{Wal:1984}, the results should be considered as illustrating samples only.

 The aim of this paper has been to introduce a new model of a dielectric charged torus encircling a charged compact object. We have proceeded by using a number of simplifying assumptions: the compact object is a 
Reissner-Nordstr\"{o}m black-hole; the torus is composed of test matter; the 
matter has a polytropic equation of state with prescribed $\Gamma$; the torus 
has constant specific angular momentum, $\ell={\rm const}$, and the specific 
charge is constant everywhere in the equatorial plane, $\gamma=1$. Despite the 
great simplifications coming from these assumptions, the scenario is still 
physically reasonable and non-trivial. Moreover, a large variety of degrees of 
freedom can be captured and the free parameters can be conveniently chosen in 
order to describe an astrophysically relevant situation. Of course, more 
complicated choices would then require more complicated calculations. Such 
calculations are now in progress, but are beyond the scope of the present 
paper.

\begin{acknowledgments}
The Opava Institute of Physics and Prague Astronomical Institute
have been operated under the projects MSM\,4781305903 and AV\,0Z10030501, and further supported by 
the Center for Theoretical Astrophysics LC06014 in the Czech Republic. JK, VK 
and ZS thank the Czech Science Foundation (ref. P209/10/P190, 205/07/0052, 
202/09/0772). We also gratefully acknowledge support from CompStar, a Research 
Networking Programme of the European Science Foundation and thank an anonymous
referee for advice and critical comments which have led to improvement of our paper.
\end{acknowledgments}

\providecommand{\noopsort}[1]{}\providecommand{\singleletter}[1]{#1}%

\end{document}